\documentclass[preprint2]{aastex}
\usepackage{graphicx}
\usepackage{dcolumn}
\usepackage{bm}

\newcommand{\ignore}[1]{}

\newcommand{\Msun}{M$_{\mbox{Sun}}$}

\newcommand{\figwidth}{3.4in}

\begin{document}


\title{Sensitivity of stellar physics to the equation of state}

\date{March 2, 2019; updates to June 28, 2020}

\author{D.C.~Swift, T.~Lockard, M.~Bethkenhagen,\footnote{%
Present address: CNRS, \'Ecole Normale Sup\'erieure de Lyon, Laboratoire de G\'eologie de Lyon, Centre Blaise Pascal,
46 all\'ee d’Italie, Lyon 69364, France
}
    A.~Kritcher, S.~Hamel, \and
   D.~Dearborn}
\affil{%
   Lawrence Livermore National Laboratory,
   7000 East Avenue, Livermore, California 94550, U.S.A.
}

\begin{abstract}
The formation and evolution of stars depends on various physical aspects
of stellar matter, including the equation of state (EOS) 
and transport properties.
Although often dismissed as `ideal gas-like' and therefore simple,
states occurring in stellar matter are dense plasmas, and the EOS has not
been established precisely.
EOS constructed using multi-physics approaches found necessary for laboratory
studies of warm dense matter give significant variations in stellar regimes,
and vary from the EOS commonly used in simulations of the formation and
evolution of stars.
We have investigated the sensitivity of such simulations 
to variations in the EOS, for sun-like and low-mass stars,
and found a strong sensitivity of the lifetime of the Sun and of the lower
luminosity limit for red dwarfs.
We also found a significant sensitivity in the lower mass limit for
red dwarfs.
Simulations of this type are also used for other purposes in astrophysics,
including the interpretation of absolute magnitude as mass,
the conversion of inferred mass distribution to the initial mass function
using predicted lifetimes,
simulations of star formation from nebul\ae, 
simulations of galactic evolution,
and the baryon census used to bound the exotic contribution to 
dark matter.
Although many of the sensitivities of stellar physics to the EOS are large, 
some of the inferred astrophysical quantities are also constrained by
independent measurements, although the constraints may be indirect and
non-trivial.
However, it may be possible to use such measurements to constrain the
EOS more than presently possible by established plasma theory.
\end{abstract}

\keywords{Equation of state - Stars: composition - Stars: interior}

\maketitle

\section{Introduction}
A unique characteristic of astrophysics is the limited information about bodies observed,
leading to a reliance on a network of supporting assumptions.
This is even true of geophysical knowledge about the Earth,
for instance because the composition of the core and mantle cannot be
measured directly, and applies even more to planets and exoplanets.
Assumptions are made about the composition and internal structure, 
using the equation of state (EOS) for relevant compositions
of matter to discriminate possible from unlikely interpretations.
The EOS even more important for stellar structure and evolution,
because, as well as guiding the interpretation of structure through
compressibility, the rate of thermonuclear reactions depends sensitively
on temperature.

We have developed laboratory experiments probing a wider ranges of states
than were accessible previously,
in particular at the National Ignition Facility (NIF)
\citep[e.g.][]{Doeppner2018,Swift2018,Kritcher2020,Swift2020}.
These experiments are able to explore
progressively more regimes of direct relevance to stellar structure.
Such measurements also provide more constraining validation of EOS models, 
where consistent physics can be used between the regimes probed experimentally
and those occurring in stars.

The purpose of work reported here is to assess the sensitivity of some key
stellar structure and evolution simulations to variations in the EOS.
We chose the lifetime of the Sun, and the lower mass limit of a red dwarf.
The latter was a proxy for lower mass limit of a brown dwarf,
which is potentially more interesting as a factor in evaluating the amount
of excess gravitational binding attributable to dark matter,
but the brown dwarfs are thought to be significantly more complicated to
simulate than red dwarfs, as discussed below.
We compare simulations of the structure and evolution of stellar bodies
with variations between representative EOS.
We also assess the relationship with uncertainty in the mass budget for exotic dark matter.

The sensitivities we present are raw, without adjusting other models.
Coupled with the uncertainty in an EOS,
the sensitivities do not imply a formal uncertainty in any particular property
of a star:
one would adjust other models such as the composition, opacity, and convective transport
within their respective uncertainties to match observables of the star and hence
re-constrain the overall model, usually reducing the effective uncertainty.
However, the raw sensitivity calculated here is the more relevant representation of the
sensitivity to a specific model such as the EOS.
Equivalent sensitivities are used widely in other applications such as inertial confinement fusion.

Analogous studies have been reported recently of the sensitivity to opacity \citep{Guzik2018}.

\section{Stellar evolution simulations}
Stellar evolution simulations are usually performed with one spatial dimension treated explicitly (the radius),
for an initial condition comprising the composition and state (mass density $\rho$ and temperature $T$)
as a function of radius.
As the star evolves, gravitational energy and thermonuclear reactions produce heat, which is conducted and convected
within the star and radiated from its surface, the composition changes via the reactions.
As the simulation usually covers a long period (billions of years) compared with the characteristic oscillation
time of the star (hours), the motion of stellar matter under pressure gradients is not tracked explicitly, 
and instead the simulation proceeds by finding the instantaneous structure that is stable with respect to gravitational buoyancy.

Simulations were performed using the computer program {\sc stars}
\citep{Eggleton1971,Eggleton1972,Eggleton1973,Eggleton1973a,Eldridge2004,Pols1995,Schroeder1997,Stancliffe2009,Stancliffe2004,Stancliffe2005,Stancliffe2007}.
For our present purposes we are interested in the calculation of material properties,
the {\tt statef} subroutine in {\sc stars}.
For comparison with alternative programs, we note the key models used for material properties.
Fermi-Dirac integrals were evaluated per \citet{Eggleton1973a}.
`High-Z' species were treated as completely ionized, which is unlikely to be accurate,
but is probably unimportant as we focused on cases with low-Z compositions.
Partition function for H$_2$ was taken from \citep{Vardya1960,Webbink1975},
representing ionization and molecular dissociation.
Molecular opacities were taken from \citep{Alexander1994},
electronic conduction from \citep{Itoh1983},
and otherwise the {\sc opal} model was used \citep{Iglesias1992}.
Dissociation coefficients were taken from \citep{RossiMaciel1983}.
Opacities for molecular CN were from \citep{ScaloUlrich1975};
CO, OH and H$_2$O were from \citep{Keeley1970,Marigo2002}.
Neutrino loss rates were taken from \citep{Itoh1983b,Itoh1989,Itoh1992}.

\begin{figure}
\begin{center}\includegraphics[width=\figwidth]{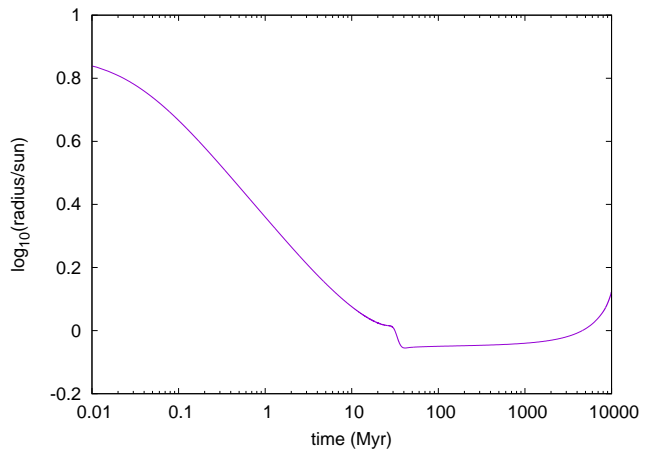}\end{center}
\caption{Evolution of a star of the mass of the Sun,
   logarithmic time scale showing early contraction.}
\label{fig:radiuslog}
\end{figure}

Each simulations began with a cloud of cool gas collapsing under gravitational attraction.
Nuclear reactions arrest the collapse and lead to a steadily-evolving star.
(Fig.~\ref{fig:radiuslog}.)

To perform a sensitivity study, the simulations were repeated with a 
perturbation made to the EOS.
Each pertubation was a scaling of the pressure and specific internal energy.
For each case, we extracted a representative state at the center, of
mass density $\rho$, temperature $T$, and composition, 
and compared the pressure calculated using different EOS.

\section{Lifetime of the Sun} 
\begin{figure}
\begin{center}\includegraphics[width=\figwidth]{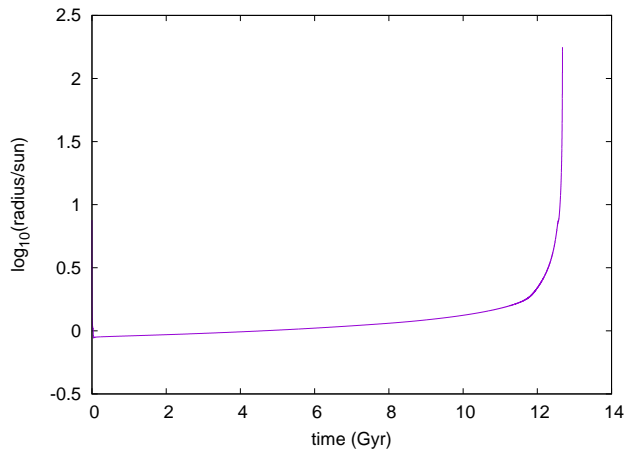}\end{center}
\caption{Evolution of the radius of the Sun in the baseline simulation.}
\label{fig:radius}
\end{figure}

\begin{figure}
\begin{center}\includegraphics[width=\figwidth]{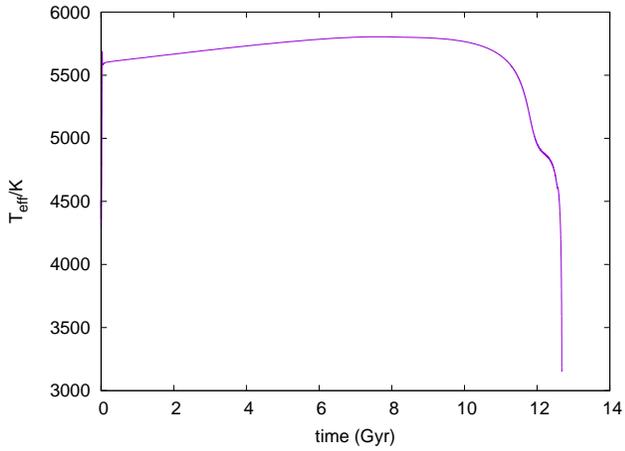}\end{center}
\caption{Evolution of the effective temperature of the Sun in the baseline simulation.}
\label{fig:teff}
\end{figure}

\begin{figure}
\begin{center}\includegraphics[width=\figwidth]{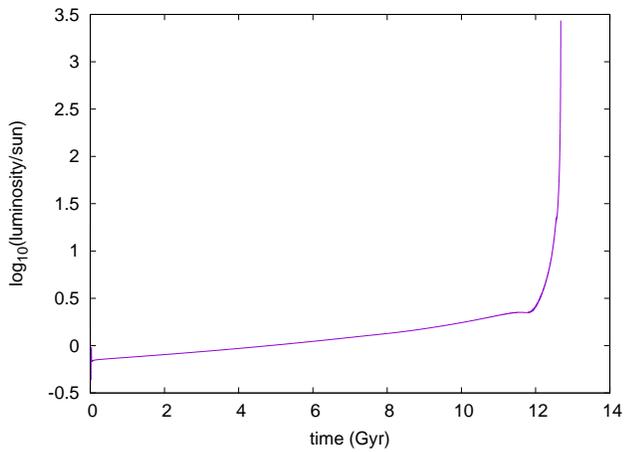}\end{center}
\caption{Evolution of the luminosity of the Sun in the baseline simulation.}
\label{fig:lumin}
\end{figure}

{\sc stars} simulations of the Sun capture its
condensation, a slow evolution through its present state, and its eventual expansion as red giant.
We performed simulations using the default solar model profile,
which has been optimized to give the currently observed radius and luminosity.
This model is consistent with a present age of $\sim$7\,Gyr,
and predicts the formation of red giant $\sim$5.5\,Gyr in the future.
(Figs~\ref{fig:radius}, \ref{fig:teff}, and \ref{fig:lumin}.)

\begin{figure}
\begin{center}\includegraphics[width=\figwidth]{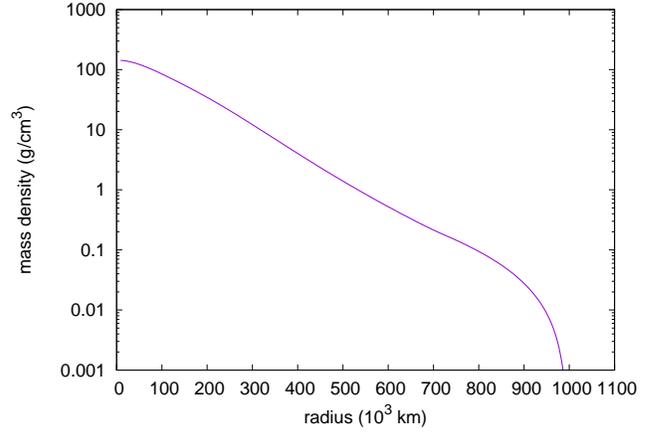}\end{center}
\caption{Mass density profile through the Sun in the baseline simulation
   at 4.6\,Gyr.}
\label{fig:dprof}
\end{figure}

\begin{figure}
\begin{center}\includegraphics[width=\figwidth]{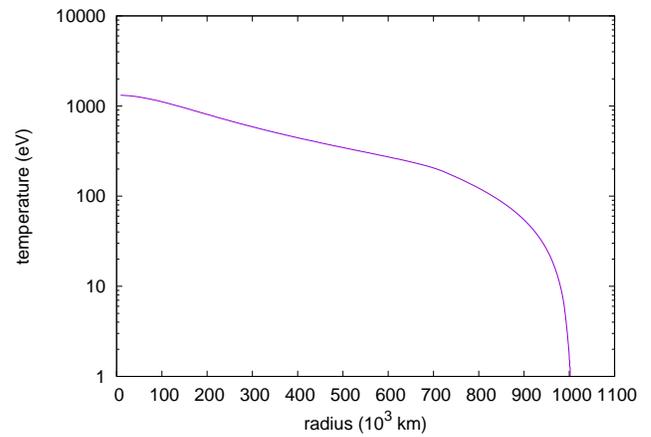}\end{center}
\caption{Temperature profile through the Sun in the baseline simulation
   at 4.6\,Gyr.}
\label{fig:tprof}
\end{figure}

\begin{figure}
\begin{center}\includegraphics[width=\figwidth]{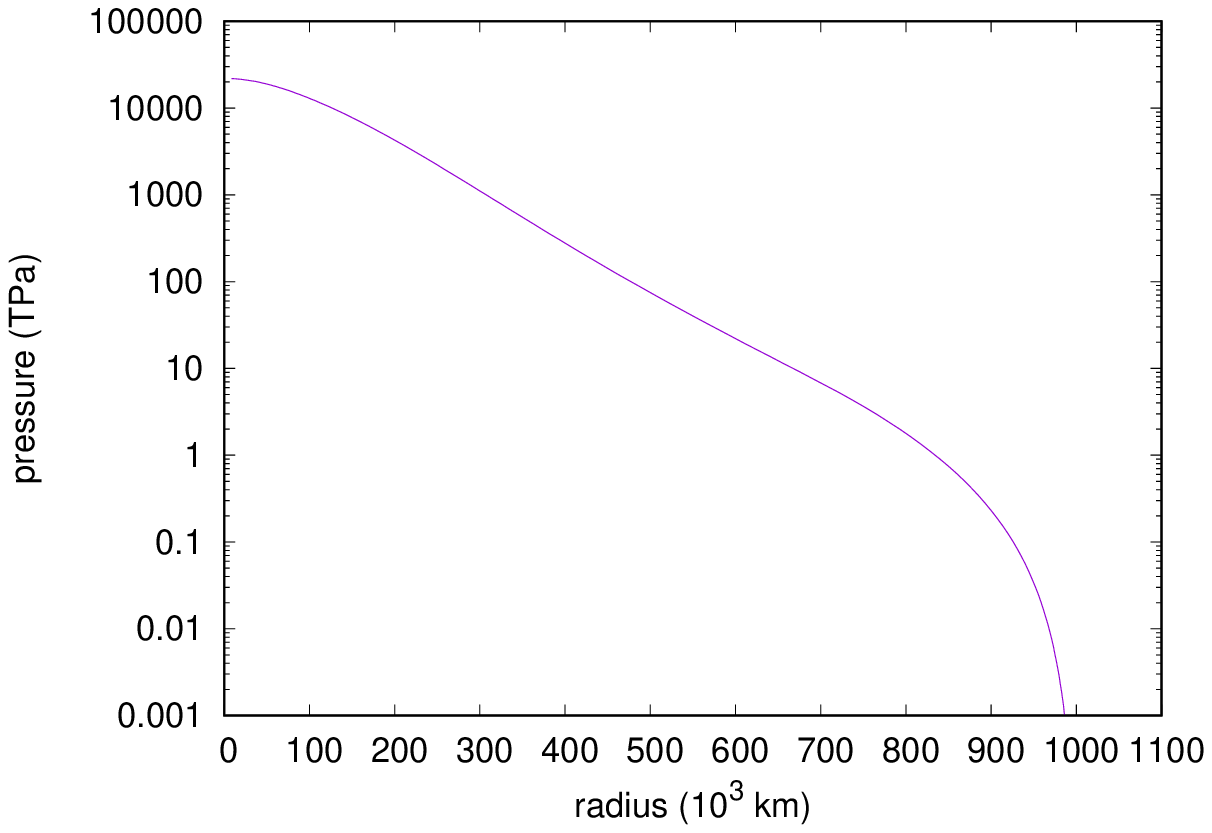}\end{center}
\caption{Pressure profile through the Sun in the baseline simulation
   at 4.6\,Gyr.}
\label{fig:pprof}
\end{figure}

It is interesting to look at the predicted structure of the Sun at the present,
in comparison with the range of data accessible to our current experimental
platforms at NIF.
(Figs~\ref{fig:dprof}, \ref{fig:tprof}, and \ref{fig:pprof}.)

\begin{figure}
\begin{center}\includegraphics[width=\figwidth]{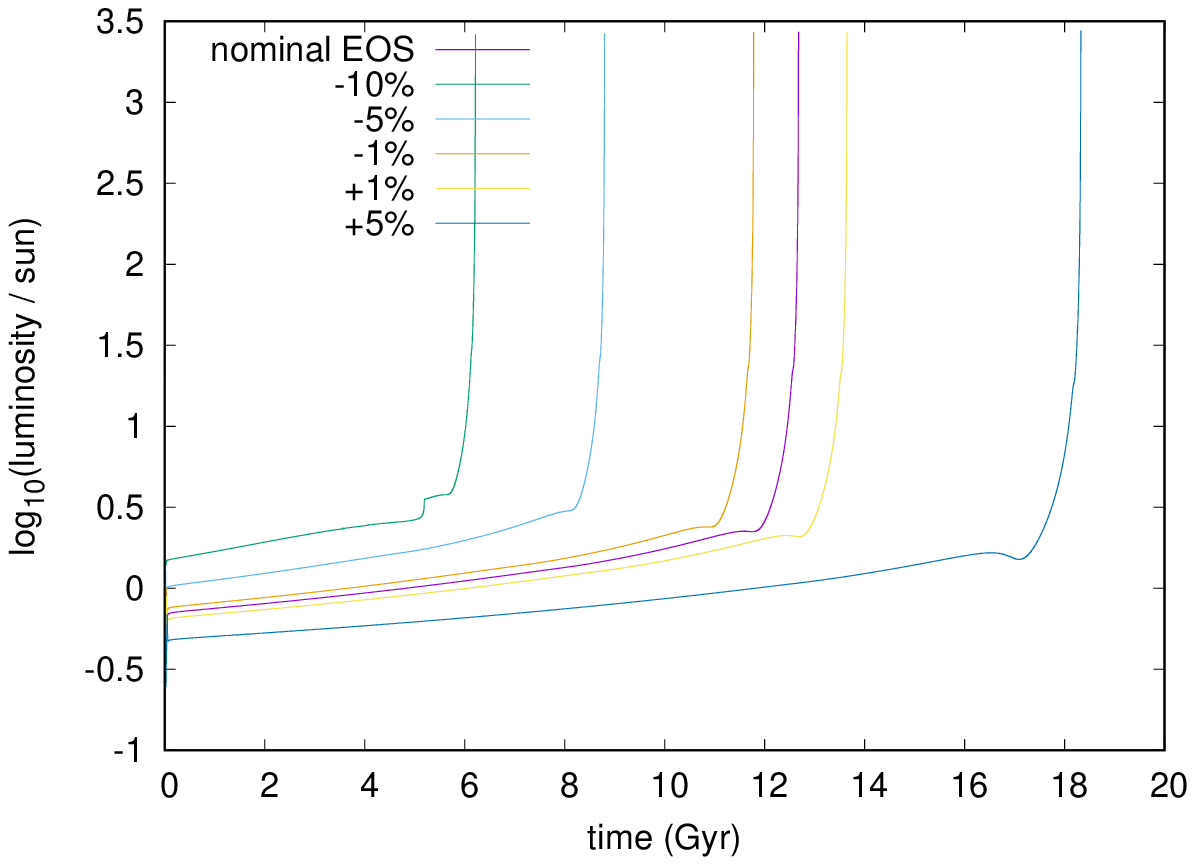}\end{center}
\caption{Sensitivity of the life of the Sun to variations in the equation of state.}
\label{fig:sunlife}
\end{figure}

Perturbations were made of a increase or decrease in stiffness of the EOS
calculation, by global constant factor.
The simulation of the evolution of the system was then repeated with the
perturbed EOS.
A percent change in the EOS produced a $\sim10$\%\ change in the lifetime,
considered as the time from formation to expansion as a red giant.
A softer EOS led to a shorter lifetime, as one would expect as the compression
at the center would then be greater and hence the rate of thermunuclear reactions
higher.
(Fig.~\ref{fig:sunlife}.)

These simulations also illustrate the significance of the raw sensitivity:
with a perturbed EOS, the radius and luminosity at $\sim 4.5$\,Gyr are 
different than observed for the Sun.
Since the age of the Sun can be estimated or constrained in other ways, 
such as by isotopic abundances, it would be necessary to compensate for
a perturbation in the EOS by adjusting other aspects of the model,
such as the elemental abundances or transport properties.
If these other properties are themselves constrained in other ways,
the simulated evolution of the Sun may be used as a constraint on the EOS,
if the underlying assumptions, observations, and chain of reasoning can be
defined precisely enough.

From the simulations, a representative state at the center of the Sun around the present
is 62\%\ by mass He and 2\%\ higher Z elements, the rest being H,
with a mass density of 162.2\,g/cm$^3$ and a temperature of 1.35\,keV.
It is interesting to compare the variations between general-purpose EOS at these conditions.
EOS are constructed for a fixed composition of matter.
Unless an EOS has been constructed for a specific composition of interest, or as a function
of composition such that the EOS for the specific composition can be obtained by interpolation,
then a mixture model is used to estimate the EOS from end-points of widely-varying composition,
such as for the component elements.
For a direct comparison of common EOS, we compare EOS for the same composition at the density
and temperature state of the center, without adjusting the mass density for composition.
EOS were taken from the Los Alamos National Laboratory {\sc sesame} library \cite{sesame1,sesame2}
and the Lawrence Livermore National Laboratory {\sc leos} library \cite{leos1,leos2},
and also from calculations using a recent development of the atom-in-jellium (AJ) method \cite{Swift_ajeos_2019},
including mixture constructions for compositions matching the {\sc stars} prediction of the
center of the Sun.
Most wide-range EOS use Thomas-Fermi (TF) theory \citep{Thomas1927,Fermi1927} for the electron-thermal energy, which
dominates under these conditions.
The AJ EOS were constructed using the {\sc inferno} computer program,
which includes alternative approaches for calculating the free energy \cite{Liberman1979,Liberman1990}.
The difference between these models gives an indication of the model uncertainty, and so we list both.
The {\sc sesame} library includes an EOS, 5280, for a composition relevant to the Sun: the
Ross-Aller solar composition.
(Table~\ref{tab:sunctreos}.)

\begin{table*}
\begin{center}
\caption{Pressure from various equations of state for the conditions at the center of the Sun.}
\label{tab:sunctreos}
\begin{tabular}{|l|l|l|}\hline
{\bf composition} & {\bf model} & {\bf pressure (PPa)} \\ \hline
H & TF: {\sc leos} 11 & 35.88 \\
  & TF: {\sc sesame} 5250, 5251 & 57.56, 42.09 \\
  & AJ & 45.15, 45.28 \\ \hline
He & TF: {\sc sesame} 5760, 5761 & 15.84, 15.93 \\
   & AJ & 15.71, 15.61 \\ \hline
H-21.4He-1.7O & TF with mixing: {\sc sesame} 5280 & 61.5 \\
H-\{62,64\}He & AJ with mixing & 27.02, 26.43 \\
H-62He-2O & AJ with mixing & 26.40 \\
H-62He-2M & {\sc stars} & 21.87 \\
\hline\end{tabular}
\end{center}
\end{table*}

It is interesting to see the difference between EOS constructed using essentially the same
widely-used model (TF) by different operators and using different computer programs, even for an element:
several tens of percent, for H.
The TF and AJ EOS for He were much more consistent, though the variations still of the order of 1\%,
which equates to a significant difference with respect to the lifetime of the Sun.
The TF-based EOS for Ross-Aller solar mixture was implausibly high in pressure.
The EOS calculation in the {\sc stars} simulation was almost 20\%\ softer than the AJ calculation,
representing the best self-consistent EOS theory used here.
Given the sensitivity of the lifetime of the Sun to the EOS, these differences are enormous,
and illustrate the degree to which uncertainties in the EOS may be masked by making adjustments to
transport calculations or thermonuclear cross-sections..

\section{Lower mass limit of a red dwarf}
As discussed below, an important question is the uncertainty in the 
mass budget for exotic dark matter attributable to the EOS.
Self-gravitating bodies can be collected into different populations,
based on the source of mass and competing mechanisms for accretion
and dispersal.
These populations include the successive generations of stars, and also
the (exo)planets that form from the residual matter following the
accretion of each star.
In each population, most mass is contained within the end of the distribution
comprising smaller bodies.
For the population of bodies accreting from stellar nebul\ae,
the small-mass bodies are the brown dwarfs.
Assessing the ratio of luminous to non-luminous mass in this population
this depends on determining the lower mass limit for a brown dwarf to
form and emit radiation by internal thermonuclear reactions.

However, brown dwarfs are thought to burn deuterium rather than hydrogen,
and so are more sensitive to the composition of the nebula from which they
form.
Brown dwarfs are also dim and relatively difficult to observe.
It is also more difficult to relate the luminosity uniquely and reliably to the mass, 
as they are thought to have dust in their atmosphere.
We consider instead the lower mass limit of red dwarfs.

\begin{figure}
\begin{center}\includegraphics[width=\figwidth]{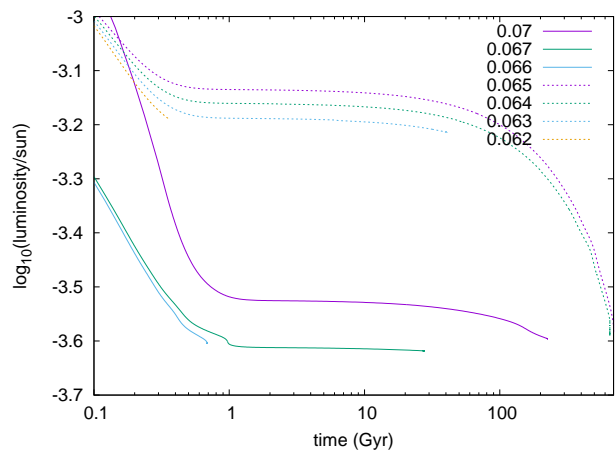}\end{center}
\caption{Sensitivity of red dwarf lighting to variations in the equation of state.
   Each curve represents a simulation with a different initial mass,
   as a fraction of that of the Sun.
   The solid lines are simulations with the baseline EOS;
   the dashed lines for EOS reduced by 10\%.}
\label{fig:redcmp}
\end{figure}

For red dwarfs, we performed simulations with different masses for the initial nebula,
and found the minimum mass at which nuclear reactions impeded the gravitational collapse
and led to a long-lived star.
This scan over masses was repeated with perturbations to the EOS,
to find the sensitivity of the minimum mass to the EOS.
With the default EOS, the lower mass limit of a red dwarf was calculated to be 0.067\,\Msun,
consistent with the accepted value.
With an EOS perturbed to be 10\%\ softer, the lower mass limit was calculated to be
0.063\,\Msun: $\sim$7\%\ less.
However, the lower luminosity limit rose from -3.6 to -3.2 (logarithm to base 10),
i.e. a factor 2.52 greater.
(Fig.~\ref{fig:redcmp}.)

From the simulations, a representative state at the center of a minimum mass red dwarf
is 28\%\ by mass He and 2\%\ higher Z elements, the rest being H,
with a mass density of 410.2\,g/cm$^3$ and a temperature of 234\,eV.
(Table~\ref{tab:reddwfctreos}.)

\begin{table*}
\begin{center}
\caption{Pressure from various equations of state for the conditions at the center of a minimum mass red dwarf.}
\label{tab:reddwfctreos}
\begin{tabular}{|l|l|l|}\hline
{\bf composition} & {\bf model} & {\bf pressure (PPa)} \\ \hline
H & TF: {\sc leos} 11 & 27.88 \\
  & TF: {\sc sesame} 5250, 5251 & 25.19, 33.86 \\
  & AJ & 22.34, 22.18 \\ \hline
He & TF: {\sc sesame} 5760, 5761 &  9.94, 10.39 \\
   & AJ &  8.22,  8.18 \\ \hline
H-21.4He-1.7O & TF with mixing: {\sc sesame} 5280 & 14.12 \\
H-\{62,64\}He & AJ with mixing & 18.42, 18.14 \\
H-62He-2O & AJ with mixing & 18.24 \\
H-62He-2M & {\sc stars} & 25.55 \\
\hline\end{tabular}
\end{center}
\end{table*}

In the cooler, denser conditions at the center of a red dwarf, the TF EOS varied over a narrower range
than for the conditions at the center of the Sun.
The difference between TF and AJ for He was an order of magnitude greater.
The EOS calculation in the {\sc stars} simulation was 40\%\ stiffer than the AJ calculation,
a striking reversal in comparison with the case of the Sun.
These differences are significant with respect to the sensitivity of the lower mass limit of a red dwarf to the EOS,
and are absolutely dominant with respect to the sensitivity of the lower luminosity limit.

\section{Mass budget for exotic dark matter} 
Dark matter was first postulated as an explanation for galactic rotation curves,
most of which imply that more mass is present at outer radii than would be implied
by the expected ratio of non-luminous matter to luminous matter in stars
(for example, \citet{Ma2014}).
Current estimates of the discrepancy suggest that exotic dark matter must
account for 85\%\ of the matter in the universe, i.e. that the ratio
of exotic dark matter to baryonic matter is $\sim 5.7$.

The ratio of non-luminous to luminous baryonic matter is estimated using models of
the formation and evolution of stars and galaxies.
The observational baryon census is non-trivial.
For stars, a model is needed to convert the absolute magnitude to mass.
This model depends on stellar structure theory, and thus on the EOS.
Observational calibrations of the relationship have significant uncertainty,
especially at low mass \citep{Armitage1996}.

As mentioned above, bodies condensing from a nebula are expected to follow a
power-law distribution which must cut off at low mass.
However, this cut-off is poorly understood;
this is important as most of the total mass occurs in low-mass objects.
In contrast, the luminosity of a galaxy is dominated by massive, short-lived stars;
these are also poorly understood.
The total baryonic mass is correlated with models of baryogenesis in the Big Bang,
and of galactic evolution.

If the mass distribution has the form $\xi_0M^{-\alpha}$,
then the ratio of total to baryonic mass varies as $M^{\alpha-1}$.
Salpeter's estimate of the mass distribution was $\alpha\simeq 2.35$ \citep{Salpeter1955}.
Using the sensitivity of red dwarf mass limit to EOS,
the uncertainty in baryonic mass is at least 30\%, 
i.e. at least 5\%\ in exotic dark matter.
Using the sensitivity of the red dwarf luminosity limit,
the uncertainty in EOS dominates, contributing tens of percent to the
uncertainty in exotic dark matter.

\section{Conclusions}
Simulations of stellar formation and evolution are very sensitive to the equation of state of stellar matter.
The lifetime of the Sun changes by $\sim$10\%\ for a 1\%\ change in the EOS.
The threshold mass for red dwarf ignition is less sensitive to the EOS, changing by $\sim$7\%\ for a 10\%\ change in the EOS,
but the luminosity at the threshold is significantly more sensitive, varying by $\sim$250\%.
The ratio of luminous to non-luminous matter in the galaxy is likely to be at least as sensitive.
Relevant EOS are uncertain at the $\sim$10\%\ level for the Sun and $\sim$30\%\ for red or brown dwarfs.
Compared to the models used in the {\sc stars} program,
the best mixture EOS currently available suggest a pressure $\sim$20\%\ greater for representative conditions at the center of the Sun,
and $\sim$25\%\ for corresponding conditions in a red dwarf.
These sensitivities are greater than the net uncertainty from observational constraints, particularly of the Sun but also of
red dwarfs,
suggesting that astrophysical observations may provide a constraint on the EOS of stellar matter as well as on models
of opacity and convection.
It would also be worthwhile investigating a wider range of stellar structure calculations using more accurate EOS models.

\section*{Acknowledgments}
This work was performed under the auspices of the U.S. Department of Energy under contract DE-AC52-07NA27344.

\end{document}